\documentclass[nofootinbib,superscriptaddress,amsmath,amssymb,aps,prb,twocolumn]{revtex4-2}
\usepackage[utf8]{inputenc}
\usepackage{graphicx}
\usepackage{dcolumn}
\usepackage{bm}
\usepackage{amssymb}
\usepackage{amsmath}
\usepackage{soul}
\usepackage{physics}
\usepackage{braket}
\usepackage[usenames, dvipsnames]{color}
\usepackage[english]{babel}
\usepackage{svg}
\usepackage[T1]{fontenc}
\usepackage{hyperref}
\usepackage{lipsum}
\usepackage{float}
\usepackage{subcaption}
\usepackage{cleveref}

\newcommand\blfootnote[1]{%
  \begingroup
  \renewcommand\thefootnote{}\footnote{#1}%
  \addtocounter{footnote}{-1}%
  \endgroup
}
\usepackage[font=small,labelfont=bf,justification=Justified]{caption}

\hypersetup{colorlinks = true, linkcolor=blue, citecolor=blue, urlcolor=blue}

\begin{document}

\title{Field-free deterministic spin-orbit torque magnetization switching using monolayer WTe$_2$}

\author{Ganesh Hanchanahal$^{\parallel}$}
\affiliation{Department of Physics, BITS-Pilani, K K Birla Goa Campus, Zuarinagar, Goa 403726, India}
\affiliation{Department of Electrical Engineering, Indian Institute of Technology Bombay, Powai, Mumbai-400076, India}

\author{Arnav Jain$^{\parallel}$}
\affiliation{Department of Physics, Indian Institute of Technology Bombay, Powai, Mumbai-400076, India}

\author{Shrushti Tapar}
\affiliation{Department of Electrical Engineering, Indian Institute of Technology Bombay, Powai, Mumbai-400076, India}
\author{Saroj P. Dash}
\affiliation{Department of Microtechnology and Nanoscience, Chalmers University of Technology, SE-41296, G\"oteborg, Sweden}

\author{Bhaskaran Muralidharan}
\email{bm@ee.iitb.ac.in}
\affiliation{Department of Electrical Engineering, Indian Institute of Technology Bombay, Powai, Mumbai-400076, India}

\begin{abstract}
An important step towards the practical implementation of spin-orbit torque magnetic memory is the field-free switching of the magnetization vector. In this work, we investigate spin-orbit torque-driven magnetization dynamics and switching devices comprising monolayer WTe$_2$ as the spin-source material. We perform a comparative analysis of the two phases of WTe$_2$, namely the 1T$_d$ and 1T$^\prime$, with the necessary symmetries for field-free deterministic magnetization switching. We demonstrate switching in both systems and analyse the dependence of the switching dynamics on the applied bias voltage, ribbon orientation and width of the magnetic layer. Modelling the ferromagnet as an exchange-coupled macrospin within a non-equilibrium Green’s function framework, we find that the x-ribbon geometry of the 1T$_d$-WTe$_2$ phase generates a significantly larger current-induced spin polarization than the 1T$^\prime$ phase, resulting in a larger damping-like torque on the adjacent magnetic layer and faster switching. The switching time of 1T$^\prime$ devices varies strongly and non-monotonically with the width of the magnetic layer. In 1T$_d$ devices, it remains nearly constant over a broad range of widths, consistent with an edge-dominated torque. These results identify x-ribbon 1T$_d$-WTe$_2$ as a promising spin source for field-free SOT-MRAM.
\end{abstract}

\maketitle

\blfootnote{$\parallel$These authors contributed equally to this work}

\section{\label{sec:level1}Introduction}

Spintronic memory technologies offer nonvolatile data storage, high endurance, and fast operation, making them attractive alternatives to conventional charge-based memories \cite{spintronics_MRAM_review,MRAMreview,Manchon,another_SOT_reviw,STT_SOT_review2}. Among these, magnetic random-access memory (MRAM) is particularly promising because information is stored in the magnetization state of the free layer of a magnetic tunnel junction (MTJ), enabling nonvolatile operation and compatibility with CMOS technologies \cite{MRAMreview,Manchon,another_SOT_reviw,STT_SOT_review2}. Within the MRAM family, spin-transfer-torque MRAM (STT-MRAM) switches the magnetization of the free layer by driving a spin-polarised current through the MTJ \cite{STTFirst, Berger1996-yb,Worledge2024-ib}. However, as the read and write operations share the same current path, the high current densities flowing through the tunnel barrier lead to reduced endurance \cite{SOT_MTJ_rev}. In contrast, spin-orbit torque MRAM (SOT-MRAM) uses an in-plane charge current injected through an adjacent spin-source layer, where spin-orbit coupling (SOC) converts it into a spin density that exerts a torque on the adjacent free ferromagnetic (FM) layer without passing through the tunnel barrier, thereby decoupling the read and write paths. This architecture reduces stress on the tunnel barrier during write operations and enables higher endurance and faster switching, making SOT-MRAM an attractive platform for high-performance magnetic memory \cite{SOtvsSTT,STT_SOT_review2,another_SOT_reviw,SOT_MTJ_rev,Miron2011-sh,Liu2012-pv}.


\begin{figure*}[htbp!]
	\includegraphics[width=\textwidth]{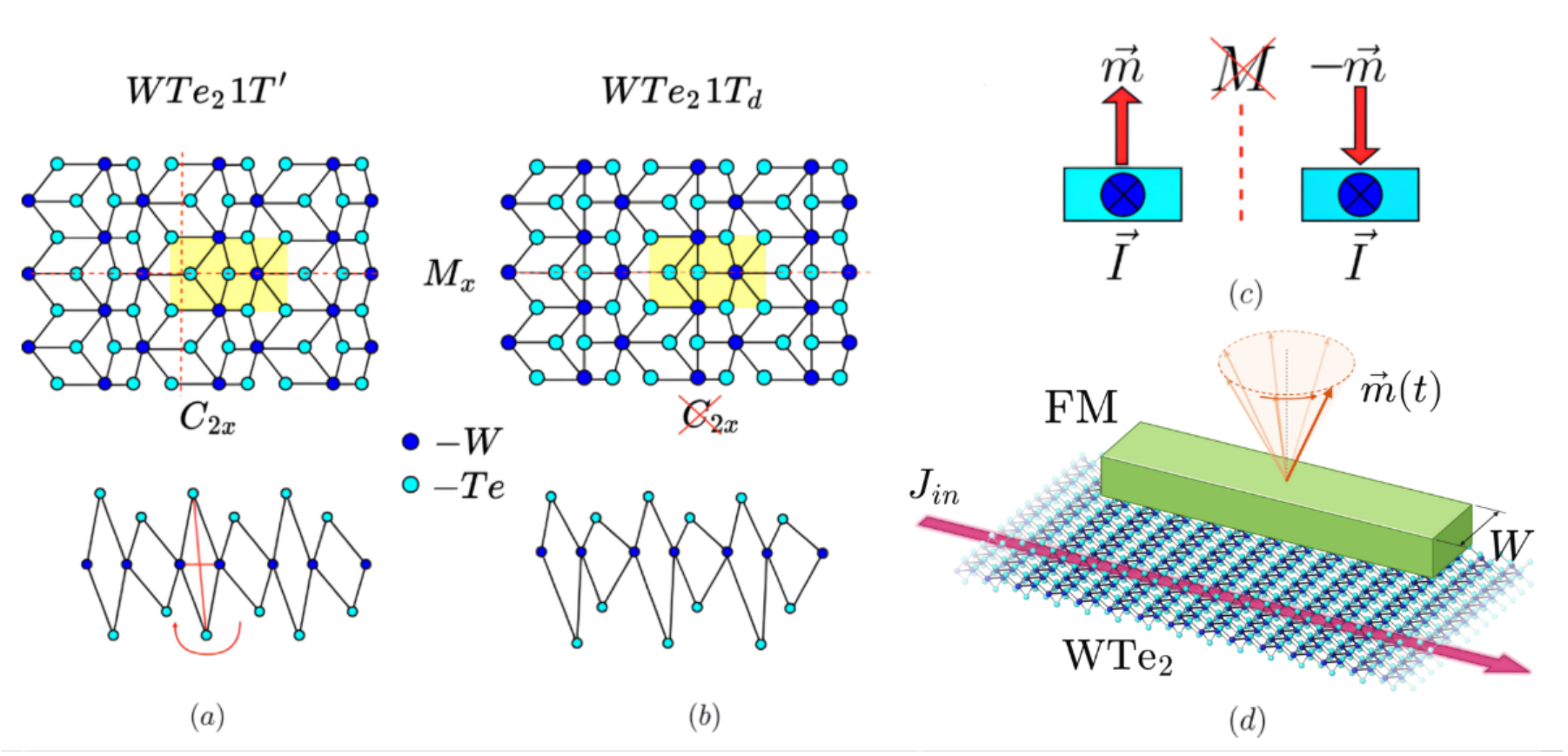}
    \caption{Schematic of the WTe$_2$ lattice structure in the 1T$^\prime$ and 1T$_d$ phases, along with the device layout and material symmetries. (a,b) Unit cells of WTe$_2$ in the 1T$^\prime$ and 1T$_d$ phases, respectively. The lower panels show side views of the lattices, where the structural distortion characteristic of the 1T$_d$ phase is evident in (b). (c) Mirror symmetry constraint for deterministic field-free SOT switching. Magnetization vector $\vec{m}$ is the stable magnetization state for current $I$. Any mirror symmetry parallel to the current flow allows for an equivalent state of opposite magnetization to be stable at the same direction of current flow. This creates two equivalently stable states, which leads to non-deterministic switching, necessitating breaking of mirror symmetries parallel to the current flow. (d) Schematic of the two-terminal device. The channel consists of monolayer WTe$_2$, which is being injected with current $J_{in}$. A ferromagnet (FM, coloured green) with width $W$ is placed on top of the WTe$_2$ layer. It is described via a macrospin approximation with magnetization $\vec{m}(t)$, which is precessing due to the spin-orbit torque applied by the spin density created in WTe$_2$.}
    \label{Fig:Fig_1}
\end{figure*} 
A central challenge in SOT-MRAM is achieving deterministic switching without an external magnetic field. Spin-orbit torques can be decomposed into two components: a field-like torque which is analogous to an effective magnetic field, and a damping-like torque which acts like an effective magnetic damping \cite{Manchon}. In conventional heavy-metal/ferromagnet bilayer systems, the spin current generated via the spin Hall effect \cite{SHE_rev} is polarised along the in-plane transverse direction, producing a damping-like torque that is symmetric with respect to both the perpendicular magnetization states; breaking this degeneracy typically requires an auxiliary in-plane magnetic field for switching of an out of plane magnet \cite{Manchon,InPlaneBField1,field_free_SOT_rev}. While several strategies have been explored to achieve field-free switching, such as interlayer exchange coupling \cite{interlayer_ex_coupling}, interlayer Dzyaloshinskii–Moriya interaction \cite{interlayer_DMI}, and vertical composition gradients in the magnetic layer \cite{vert_gradient}, these approaches introduce additional material or structural complexity. A more intrinsic route is to exploit crystal symmetry, as a spin-source layer with broken inversion and appropriate mirror symmetries can generate an out-of-plane canted spin polarization, producing a torque on the FM layer that lifts the degeneracy between the two magnetization states without an external magnetic field \cite{IEEE_symmetry,field_free_SOT_rev,2DSOT,demo_SOT_WTe2,Kang2025-ed,SP1,SP2}.

The monolayer WTe$_2$ is a 2D-transition-metal dichalcogenide (TMD) that is particularly well-suited to satisfy the symmetry requirements outlined above. Unlike conventional 2H or 1T TMDs \cite{2H_phase_stability_claim}, monolayer WTe$_2$ is the most stable in the 1T$^\prime$ phase \cite{Qian_2014,Duerloo2014-cw}. The 1T$_d$ phase differs from the 1T$^\prime$ by breaking the inversion symmetry, as shown in Fig.~\ref{Fig:Fig_1}(a,b) \cite{1Td_theory}. Both phases of monolayer WTe$_2$ are theoretically expected to be quantum spin Hall insulators with edge states exhibiting canted out-of-plane spin-polarization \cite{1Td_theory,1Tp_akhmerov}, enabling out-of-plane damping-like torques on the FM layer \cite{demo_SOT_WTe2}. While 1T$^\prime$ has been studied quite extensively experimentally \cite{1Tp_experimental, 1Td_experimental, Fei_2017,Wu_2018, Jia2017-tt}, the monolayer of 1T$_d$ phase has been recently synthesised \cite{Isolating1Td}. The unconventional out-of-plane spin-orbit torque in multilayer WTe$_2$ / FM bilayers was first experimentally demonstrated by MacNeill et al. \cite{demo_SOT_WTe2}, illustrating the utility of WTe$_2$ for crystal-symmetry engineered SOT. Subsequent experiments have demonstrated field-free deterministic magnetization switching in a range of WTe$_2$-based heterostructures, including systems employing unconventional spin-orbit and orbital torques \cite{WTe2_1,WTe2_2,WTe2_3,WTe2_4,WTe2_5,WTe2_6}. However, these works primarily focus on multilayer or few-layer devices. A direct phase comparison of 1T$_d$ and 1T$^\prime$ phases of monolayer WTe$_2$, to the best of our knowledge has not been reported, motivating the present theoretical study.


In this work, we simulate the SOT generation and magnetization dynamics of the FM for both 1T$_d$ and 1T$^\prime$ phases of WTe$_2$ and perform a comparative study between the two phases. We find that both phases apply SOT on the FM layer and show deterministic field-free switching. By further analysing the dependence of the switching time on the applied bias voltage and width of the magnetic layer, we conclude that the 1T$_d$ phase in the x-ribbon geometry is better suited for this application, owing to the larger SOT it exerts on the FM layer, its shorter switching times and weaker sensitivity to the magnet width. The paper is organised as follows: Section \ref{Sec II} presents the theoretical framework underlying the symmetry constraints and the Non-Equilibrium Green's Function (NEGF) formalism combined with the Landau-Lifshitz-Gilbert (LLG) equation to compute the spin-orbit torque on the FM layer and the resulting magnetization dynamics, Section \ref{Sec3} presents the results and compares the performance between the two phases, Section \ref{Sec4} concludes the results and summarises future work that can be done for a more realistic simulation of a SOT device.


\section{Theory}\label{Sec II}

In this section, we outline the theoretical framework used to compute SOT and the resulting magnetization dynamics in monolayer WTe$_2$-based devices. Our approach combines a symmetry-consistent tight-binding description of the electronic structure with the NEGF formalism \cite{NEGF} to evaluate SOT. The time evolution of the adjacent FM layer under the influence of the SOT is then obtained by solving LLG equation. We first discuss the symmetry requirements necessary for deterministic, field-free switching, followed by a description of the material Hamiltonians and transport formalism, and finally outline the procedure used to compute the magnetization dynamics\cite{Soleimani,Nikolic}.

\subsection{Symmetry requirements}

The generation of SOT and the realisation of deterministic, field-free magnetization switching are constrained by the inversion and mirror symmetries of the device as follows \cite{IEEE_symmetry}: 

\begin{enumerate}
    \item \textit{Inversion symmetry}: 
    Broken inversion symmetry is a necessary condition for obtaining a non-vanishing SOT. In the absence of inversion symmetry, an applied electric field can generate a non-equilibrium spin density via the Rashba-Edelstein effect, which acts as the microscopic source of SOT. If inversion symmetry is preserved, these contributions cancel, resulting in a vanishing net torque. 

    \item \textit{Mirror symmetry}:
    magnetization is an axial vector, and mirror symmetries impose additional constraints on current-induced switching. If a mirror plane parallel to the current direction is preserved, the system admits two symmetry-related magnetization configurations that are equivalent for a given current direction as shown in Fig.~\ref{Fig:Fig_1}c, leading to non-deterministic switching. Deterministic field-free switching therefore requires that the mirror symmetry parallel to the current flow be broken, while a mirror symmetry perpendicular to the current direction may be preserved.
\end{enumerate}

These symmetry considerations make monolayer WTe$_2$ a particularly suitable platform for realising deterministic, field-free SOT switching. As outlined in Fig.~\ref{Fig:Fig_1}a, in the 1T$^\prime$ phase, the crystal structure preserves a glide mirror symmetry ($M_x$) and a screw rotation symmetry ($C_{2x}$), which together give rise to inversion symmetry \cite{WTe2_1Tp_symmetry}. In contrast, the distorted $1T_d$ phase breaks $C_{2x}$ symmetry, which leads to inversion symmetry breaking. Hence 1T$_d$ automatically satisfies the necessary conditions stated above \cite{1Td_theory}. While 1T$^\prime$ does not, one can break the local inversion symmetry by placing an adjacent FM layer, enabling field-free SOT generation. In both phases, appropriate device geometries can be chosen to break the mirror symmetry parallel to the current direction (or it is already broken for x-ribbon geometries), thereby allowing deterministic switching without the need for an external magnetic field.

\subsection{Material Hamiltonian and device model}


Both the phases of monolayer WTe$_2$ are quantum spin Hall insulators, exhibiting canted spin polarization along with strong SOC. To capture these properties, we employ symmetry-consistent tight-binding Hamiltonians derived in previous works for each phase. For the 1T$_d$ phase, we use a model given by JH Garcia et al, which was constructed from symmetry-allowed terms and fitted to DFT calculations \cite{1Td_theory}, while for the 1T$^\prime$ phase we adopt a tight-binding description obtained from a combination of DFT, symmetry analysis, and ARPES data derived by A Lau et al \cite{1Tp_akhmerov}. The Hamiltonian and self-energies for the 1T$^\prime$ phase were calculated using the quantum transport software package \texttt{Kwant} \cite{Kwant}. These Hamiltonians provide an accurate description of the low-energy band structure relevant for our work. However, we need to note that these Hamiltonians are only valid for small biases.

The device geometry considered throughout this work is shown in Fig.~\ref{Fig:Fig_1}d. It consists of a finite WTe$_2$ scattering region connected to semi-infinite left and right leads and subjected to a bias voltage, and a FM layer on top which is approximated as a macrospin $\vec{m}$. The FM layer is exchange-coupled to selected lattice sites of the WTe$_2$ channel via a surface exchange interaction,
\begin{equation}
    U(t) = -\frac{\Delta_{\mathrm{surf}}}{2}\,\vec{m}(t)\cdot\boldsymbol{\sigma},
\end{equation}
where $\Delta_{\mathrm{surf}}$ denotes the exchange strength, $\vec{m}(t)$ is the time-dependent unit magnetization vector of the FM, and $\boldsymbol{\sigma}$ is the vector of Pauli matrices. The total Hamiltonian of the device is then given by $H(t) = H_0 + U(t)$, where $H_0$ is the tight-binding Hamiltonian of the isolated WTe$_2$ monolayer. This coupling locally breaks inversion symmetry in the 1T$^\prime$ phase enabling current-induced SOTs in devices of both phases. We infer the relative edge and bulk contributions to the torque from the dependence of the switching time on magnet width (\ref{Sec3}C)

\begin{figure*}[htbp!]
    \includegraphics[width=\linewidth]{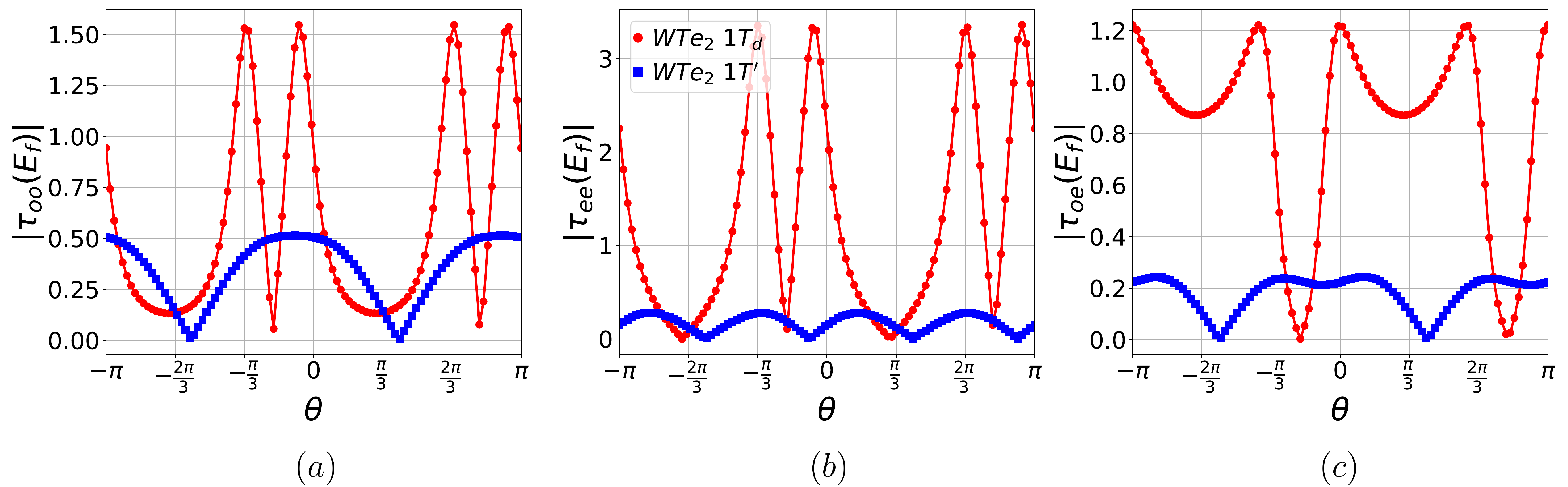}
    \caption{Magnitude of the torque components as a function of the polar angle $\theta$, with the azimuthal angle fixed at $\phi = \frac{\pi}{2}$ for WTe$_2$ 1T$_d$ and 1T$^\prime$ x-ribbon geometries. The scattering region of $16\times16$ unit cells, and the magnetic layer is placed on the top edge only. The three panels correspond to (a) $|\tau_{oo}|$, (b) $|\tau_{ee}|$, and (c) $|\tau_{oe}|$, where $\tau_{ee}$ and $\tau_{oo}$ represent the equilibrium and non-equilibrium field-like components, respectively, and $\tau_{oe}$ denotes the damping-like component of the SOT. The $\tau_{eo}$ can be shown to vanish identically \cite{STT_SOT_rev_Nikolic}, and hence not shown here. Over the majority of the angular range, the $(\text{1T}_d)_x$ /FM device exhibits a larger torque magnitude on the FM layer compared to the $(\text{1T}^\prime)_x$/FM for all components.}
    \label{fig:Fig_2}
\end{figure*}


\begin{figure*}[htbp!]
    \centering
    \includegraphics[width=\linewidth]{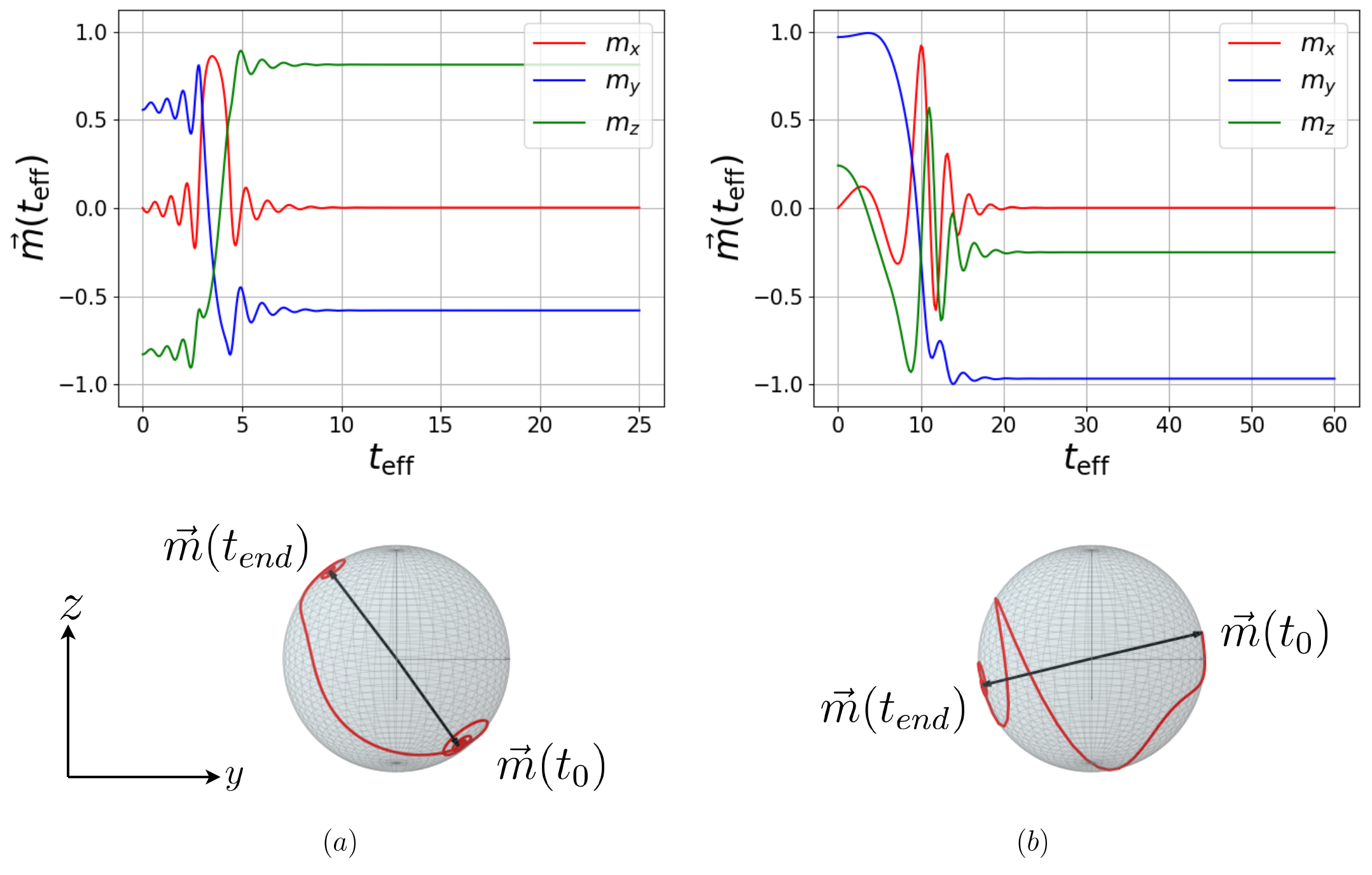}
    \caption{Magnetization dynamics using a monolayer WTe$_2$ 1T$_d$ and 1T$^\prime$ x-ribbon geometries, with the scattering region of size $16\times 16$ unit cells. A bias of $V_{\text{eff}} = 10$ is applied to both the devices. The magnet is kept on only the topmost row of unit cells (a) The spin polarization vector for $1T_{d}$ x-ribbon is $(0, -\cos(54.5^\circ), \sin(54.5^\circ))$. The initial condition is slightly canted away from $-\vec{\sigma}_P$, showing clean switching to $\vec{\sigma}_P$. (b) The spin polarization vector for 1T$^\prime$ x-ribbon is $(0,-\cos(14.5 ^\circ),-\sin(14.5^\circ))$. Here also the initial condition is slightly canted from $-\vec{\sigma}_P$ and shows switching to $\vec{\sigma}_P$. For both (a) and (b) the trajectory of the magnetization is shown on a Bloch sphere in the bottom panels.}
    \label{fig:Fig_3}
\end{figure*}

\subsection{Transport formalism and magnetization dynamics}

The SOTs acting on the magnetic layer are computed using the NEGF formalism applied to the time-dependent device Hamiltonian $H(t)$ \cite{Nikolic,Soleimani}. For a given magnetization configuration $\vec{m}(t)$, the retarded Green’s function of the scattering region is given by\cite{NEGF}
\begin{equation}
    G(E,t) = \left[E - H(t) - \Sigma_L(E) - \Sigma_R(E)\right]^{-1},
\end{equation}
where $\Sigma_{L,R}$ denote the self-energies of the left and right leads. The applied bias voltage enters through the difference in the electrochemical potentials of the leads. Throughout this work, the FM layer is treated as an exchange-coupled perturbation and not as an additional transport channel, hence no current is allowed to flow through the FM.

The bias-induced nonequilibrium spin density is obtained from the lesser Green’s function and serves as the microscopic origin of the SOT. The resulting torque acting on the magnetization is evaluated following established NEGF-based formulations and can be decomposed into symmetry-distinct components, including field-like and damping-like contributions. In equilibrium, the torque component is found to be aligned with the intrinsic spin-polarization axis of edge states in WTe$_2$, and hence can be absorbed into the anisotropy torque $\vec{T}_{MCA}$\cite{Soleimani,Nikolic}. The energy resolved torque components are $\vec{\tau}_{\mu,\nu}(E)$, where $\mu,\nu$ denote whether the torque component is odd ($o$) or even $(e)$ under bias-reversal and time-reversal respectively.

The time evolution of the magnetization is obtained by solving the LLG equation,
\begin{equation}
\frac{d\vec{m}}{dt}
= \alpha \left( \vec{m} \times \frac{d\vec{m}}{dt} \right)
+ \vec{T}_{\mathrm{MCA}}
+ \frac{\gamma }{\mu}\vec{T}_{\mathrm{SOT}}
- \gamma (\vec{m} \times \vec{B})
\end{equation}

where $\gamma$ is the gyromagnetic ratio, $\mu$ is the magnetic moment, $\mathbf{B}$ is the external magnetic field, $\vec{T}_{\text{MCA}} = \Delta_{\text{MCA}}(\vec{m}\cdot\vec{\sigma_P})(\vec{m}\times\vec{\sigma_P})$ is the magnetic anisotropy aligned with the spin polarization vector $\vec{\sigma_P}$, $\alpha$ is the Gilbert damping parameter and $\vec{T}_{SOT}$ is the SOT produced. We can simplify this equation by defining dimensionless time $t_{\text{eff}} = t\Delta_{\text{MCA}}$ and bias voltage $V_{\text{eff}}=\gamma eV_b/(\mu\Delta_{\text{MCA}})$, which will be used to perform simulations for the results.

The magnetization dynamics are simulated self-consistently by iterating between the SOT calculation and the LLG equation. Starting from an initial magnetization configuration, the torque is evaluated for the corresponding $H(t)$, the magnetization is updated over a small time step, and the procedure is repeated until the magnetization converges to a steady state or switching is achieved. This approach allows to incorporate time-dependence of the SOT, giving a more accurate description of the dynamics involved. The full details of the theory are laid out in Appendix \ref{appendix}, and the physical constants used are give in \ref{Sec3}B.

 \section{Results and Discussion} \label{Sec3}

In this section, we present numerical results for SOT generation and magnetization dynamics in monolayer WTe$_2$-based SOT devices at zero magnetic field, focusing on the contrasting behaviour of the 1T$^\prime$ and 1T$_d$ phases. Using NEGF-based torque calculations combined with Landau--Lifshitz--Gilbert dynamics, we first demonstrate deterministic, field-free magnetization switching in both systems and compare the magnitude and angular dependence of the torque components. All the results have been calculated by using dimensionless time and bias voltage, with $\Delta_{\text{surf}} = 0.1$ eV and $\alpha = 0.01$. We then analyse the dependence of the switching time on the applied bias voltage and ribbon orientation, identifying clear performance trends and voltage thresholds across different device geometries. Finally, we investigate the role of magnetic layer width and show that switching time in 1T$_d$ devices remains constant to variations in magnet width. It increases only when the magnet covers both the edges of the WTe$_2$ layer, where the torque from the opposite edge partly cancels that from the first. The 1T$^\prime$ devices exhibit strong and non-monotonic sensitivity to variations in the magnet width due to enhanced bulk contributions. Together, these results establish 1T$_d$-WTe$_2$ x-ribbon geometries as the most favourable platform for fast and reliable field-free SOT switching.

\subsection{Magnetization Switching}

We first examine the angular dependence and magnitude of the spin-orbit torque acting on the magnetic layer in the WTe$_2$ devices for the 1T$_d$ and 1T$^\prime$ phases. Fig.~\ref{fig:Fig_2} shows the energy-resolved torque components as a function of the polar angle $\theta$ of the magnetization for x-ribbon geometries, with the azimuthal angle fixed at $\phi=\pi/2$. All torque components exhibit a strong dependence on the magnetization direction, reflecting the underlying spin texture of WTe$_2$. Over most of the angular range, the 1T$_d$ x-ribbon produces a larger torque magnitude on the magnetic layer than its 1T$^\prime$ counterpart, particularly for the damping-like component $\tau_{oe}$, which is expected to be primarily responsible for driving magnetization reversal. This enhancement is consistent with the intrinsically broken inversion symmetry of the 1T$_d$ phase, whereas in the 1T$^\prime$ phase inversion symmetry is broken only through the presence of the FM layer.

The resulting magnetization dynamics, obtained by solving the Landau-Lifshitz-Gilbert equation using the self-consistently calculated torques, are shown in Fig.~\ref{fig:Fig_3}. For both phases, an initial magnetization slightly canted away from the spin-polarization direction undergoes deterministic, field-free switching to the opposite stable orientation upon application of a bias voltage. The stable orientation is different for the two phases: $\vec{\sigma_P}_{,(\text{1T}_d)} = (0, -\cos(54.5^\circ), \sin(54.5^\circ))$ for 1T$_d$ and $\vec{\sigma_P}_{,(\text{1T}^\prime)} = (0,-\cos(14.5 ^\circ),-\sin(14.5^\circ)) $ for 1T$^\prime$. Consistent with the larger torque magnitudes, the switching time in the 1T$_d$ x-ribbon is significantly shorter than in the 1T$^\prime$ x-ribbon for comparable device parameters, making it a better candidate for potential SOT applications.

\begin{figure*}[htbp!]
    \centering
    \includegraphics[width=\linewidth]{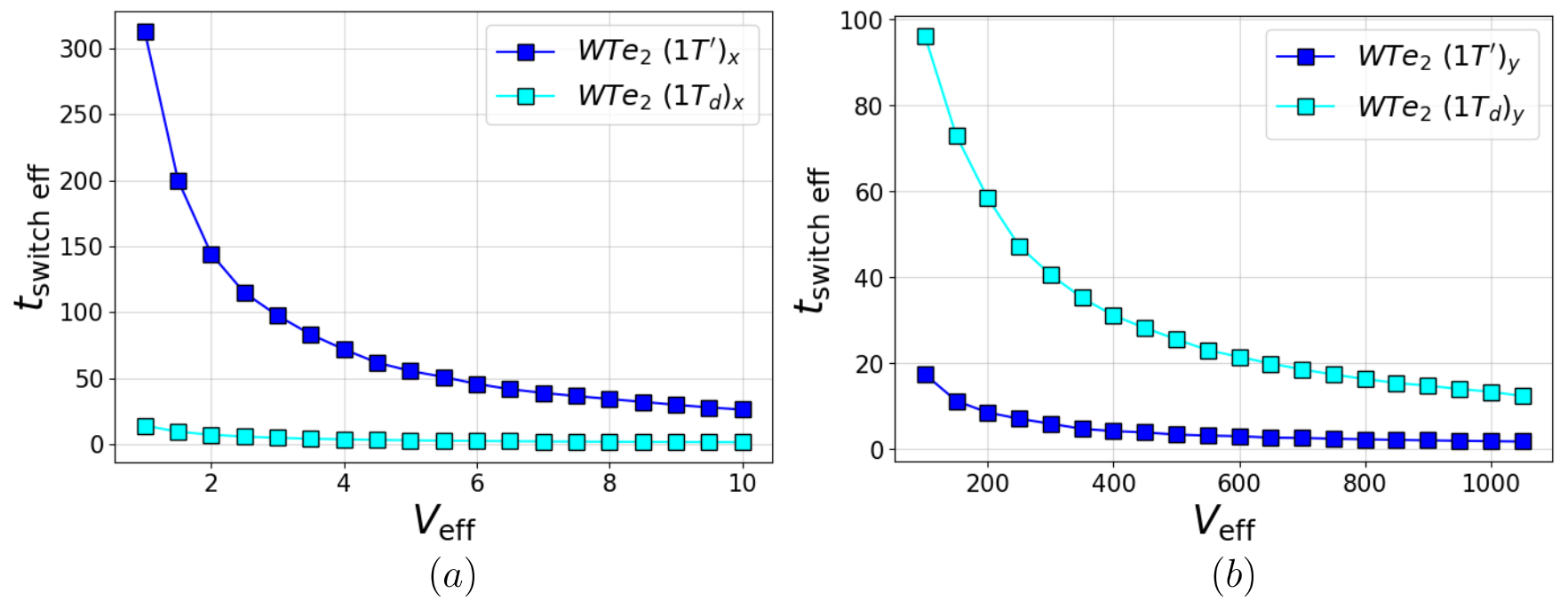}
    \caption{Switching time as a function of bias voltage for (a) x-ribbons with scattering region of size $16\times16$ unit cells and (b) y-ribbons of WTe$_2$ in the 1T$^\prime$ and 1T$_d$ phases with $12\times12$ unit cells. The magnet is placed only on the top row of unit cells for all devices. Among all cases, the $(\text{1T}_d)_x$ ribbon exhibits the fastest switching, followed by $(\text{1T}^\prime)_x$, $(\text{1T}^\prime)_y$, and $(\text{1T}_d)_y$. Notably, both y-ribbons require significantly higher minimum bias voltages (nearly two orders of magnitude larger) for switching compared to the x-ribbons.}
    \label{fig:Fig_4}
\end{figure*}

\subsection{Voltage and ribbon-geometry dependence}

We next examine how the magnetization switching time depends on the applied bias voltage and ribbon orientation for monolayer WTe$_2$ devices in the 1T$_d$ and 1T$^\prime$ phases. Fig.~\ref{fig:Fig_4} shows the switching time as a function of bias voltage for both x- and y-ribbon geometries. In all cases, increasing the applied voltage leads to faster switching, reflecting the monotonic increase in the magnitude of the non-equilibrium spin-orbit torque with bias. However, the improvement in switching time exhibits diminishing returns at higher voltages. Note that our analysis uses the small-bias approximation, and hence cannot accurately predict the behaviour at large bias voltage.

A clear distinction in device performance emerges from the comparison across phases and geometries. Among all configurations studied, the 1T$_d$ x-ribbon exhibits the fastest switching, followed by the 1T$^\prime$ x-ribbon, while both y-ribbon geometries require substantially higher bias voltages to achieve switching. In particular, the minimum voltage required for switching in y-ribbons is nearly two orders of magnitude larger than that for x-ribbons in both phases. However, as the device sizes are different, this is a qualitative comparison. The switching time at $V_{\text{eff}}=10$ in the x-ribbon orientation for 1T$_d$ is $t_{\text{eff}}=1.55$, while for 1T$^\prime$ $t_{\text{eff}}=26.29$, which is approximately 17 times that of 1T$_d$. For $\gamma = 1.76\times10^{11}\ \text{s}^{-1}\text{T}^{-1}$, magnetic moment $\mu = \mathcal{O}(\mu_B) \sim 9.274\times10^{-24} \ \text{J}\ \text{T}^{-1}$, $e = 1.6\times10^{-19} C$ and $\Delta_{\text{MCA}}=\mathcal{O}(\gamma H_K)\sim10^{11}\ s^{-1}$ (an anisotropy field $H_K\approx 0.6\ \text{T}$), we get a value of $t \sim 15.5\ \text{ps}$ for 1T$_d$ at $V_b=0.33\ \text{mV}$. These estimates are order-of-magnitude only due to the qualitative nature of this study and lie within the small-bias regime. The better performance of the 1T$_d$ x-ribbon can be traced to its larger damping-like torque over a broad range of magnetization angles because it satisfies all the symmetry requirements intrinsically, whereas in 1T$^\prime$ the inversion symmetry is broken only on addition of the magnet on top. Hence we can conclude that x-oriented 1T$_d$-WTe$_2$ devices are more favourable for field-free switching.



\subsection{Magnetic layer size dependence}

\begin{figure*}[htbp!]
    \centering
    \includegraphics[width=\linewidth]{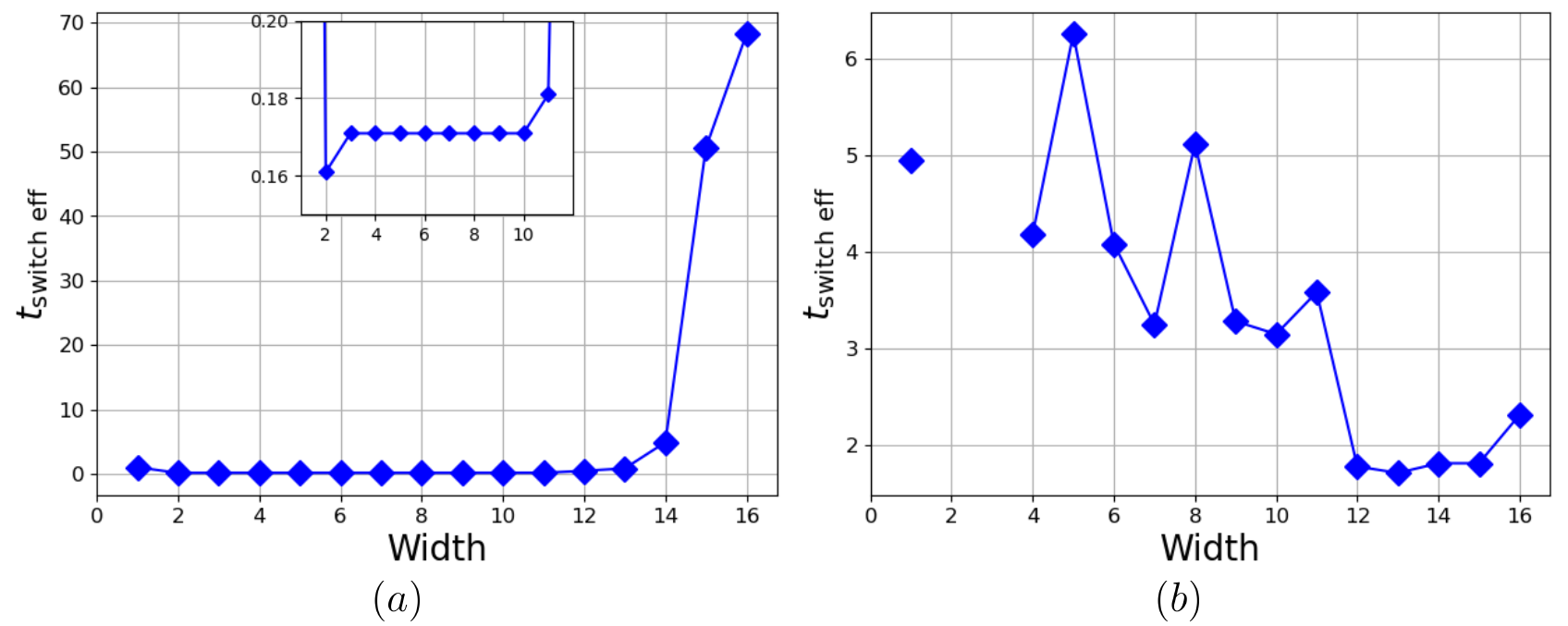}
    \caption{Switching time as a function of magnet width for WTe$_2$ in (a) the $(\text{1T}_d)_x$ phase and (b) the $(\text{1T}^\prime)_x$ phase devices with scattering region of $16\times16$ unit cells at $V_{\text{eff}}=100$. The magnet width is varied while the length is kept constant spanning the entire channel. For $(\text{1T}_d)_x$, the switching time remains nearly unchanged with increasing width, indicating that the torque originates primarily from the edges with negligible bulk contribution. It is therefore not advisable to place the magnet over the entire device, since the spin polarization at the two edges is opposite; the torque arising from the bottom edge opposes that from the top edge (though with unequal magnitude), resulting in longer switching times. In contrast, for $(\text{1T}^\prime)_x$, the switching time varies non-monotonically with width, reflecting enhanced torque generation due to bulk contributions. Notably, no switching occurs in the 1T$^\prime$ phase for magnets two and three unit cells wide.}
    \label{fig:Fig_5}
\end{figure*}

We finally investigate the dependence of the switching dynamics on the width of the magnetic layer, which is an important consideration for realistic device fabrication. In all previous simulations, the magnetic layer was restricted to a single unit cell in width and extended along the full length of the channel, thereby breaking the inversion symmetry along with the mirror plane parallel to the current flow. To assess the robustness of the switching behaviour, we systematically vary the magnet width while keeping its length fixed and compute the corresponding switching times, as shown in Fig.~\ref{fig:Fig_5}. For the 1T$_d$ x-ribbon, the switching time decreases slightly as the magnet width increases from one to two unit cells and then remains nearly constant over a broad range of widths before increasing sharply for very wide magnets. This behaviour indicates that the dominant contribution to the spin-orbit torque arises from edge states, with negligible bulk contribution over most of the device width. At the other edge, the switching time shoots up, indicating incomplete cancellation of SOT from the other edge. However, the SOT does not vanish due to inversion symmetry breaking making the two edges .

In contrast, the 1T$^\prime$ x-ribbon exhibits a markedly different and non-monotonic dependence on magnet width. Switching occurs for a one unit cell wide magnet, but not for 2 and 3 unit cells. Beyond that, the switching time varies irregularly with width, decreasing overall towards the full channel width. This behaviour suggests a substantial contribution from bulk states, possibly due to a smaller band gap. The pronounced sensitivity of 1T$^\prime$ devices to magnet size makes them less robust compared to the 1T$_d$ devices.

\section{Conclusion and Outlook} \label{Sec4}
Topological insulators are attractive platforms for nanoscale devices due to their robustness against disorder and noise. Previous studies have explored their potential in magnetic memory applications, including proposals based on the Kane–Mele model \cite{Soleimani} and experimentally realized topological insulators \cite{Nikolic}. However, these works suffer from key limitations: the Kane–Mele model possesses an unrealistically small band gap, making experimental realisation challenging, while other material platforms, despite being experimentally accessible, fail to satisfy the symmetry requirements necessary for field-free deterministic magnetization switching. In this work, we overcome both limitations by identifying a material system that is experimentally realizable and simultaneously satisfies the required symmetry constraints. Furthermore, we provide a more comprehensive analysis of the conditions necessary to achieve reliable magnetization switching.

We performed a comparative study of current-induced magnetization switching in the 1T$_d$ and 1T$^\prime$ phases of WTe$_2$. Using symmetry arguments, we demonstrated that deterministic switching can be achieved in the absence of external magnetic fields. The superior switching performance of the 1T$_d$ phase in the x-ribbon geometry correlated with the larger damping-like torque it exerts on the FM layer over a broad range of magnetisation angles, which we attribute to its intrinsically broken inversion symmetry. We found that, in both phases, the x-ribbon geometry requires a significantly lower minimum voltage to initiate switching. Additionally, independent of the material phase or device geometry, the improvement in switching time saturates with increasing applied voltage. We also observed that the switching time in the 1T$_d$ phase exhibits weaker sensitivity to variations in magnet width, owing to the dominant contribution of edge-induced torque. In contrast, for the 1T$^\prime$ phase, comparable edge and bulk torque contributions result in a stronger dependence of the switching dynamics on the magnet width.

Our results rely on several simplifications that future work should relax. The FM is modelled as an insulating macrospin coupled only through exchange, neglecting any current passing through it. Since the edge-dominated torque depends on a clean bulk gap, edge disorder, thermal broadening and larger device sizes should be examined before the width robustness of 1T$_d$ devices can be generalized. The influence of strain \cite{StrainInWTe2} on the switching efficiency and overall device performance in WTe$_2$ is another open question. Additionally, the use of counter-propagating spin currents in quantum spin Hall insulators offers an intriguing avenue for extending these concepts to the deterministic switching of antiferromagnetic systems.


\begin{acknowledgments}

We thank Dr. Venkatesh Vadde for his valuable insights and helpful discussions, particularly during the initial setup of the simulation framework used in this work. The author BM acknowledges funding from the Department of Science and Technology (DST), Government of India, under the National Quantum Mission, {through Grant no. DST/QTC/NQM/QMD/2024/4}, the Institute of Eminence Grant from IIT Bombay, {through Grant no. RD/0523-IOE00I0-139} and the Inani Chair Professorship fund{, through Grant No. DO/2024-INAN/001-001}. SPD acknowledges funding from Swedish Research Council VR (No. 2025-03702). The authors BM and ST acknowledge funding from the Dhananjay Joshi Endowment award from IIT Bombay{, through Grant No: DO/2023-DJEF002}.

\end{acknowledgments}

\textbf{Data availability}
The data that support the findings of this study are available from the corresponding author upon reasonable request.\\

\appendix

\onecolumngrid
\section{Methodology}\label{appendix}
\begin{figure*}[!h]
    \centering
    \begin{subfigure}{0.48\textwidth}
    \centering
    \includegraphics[width=\textwidth]{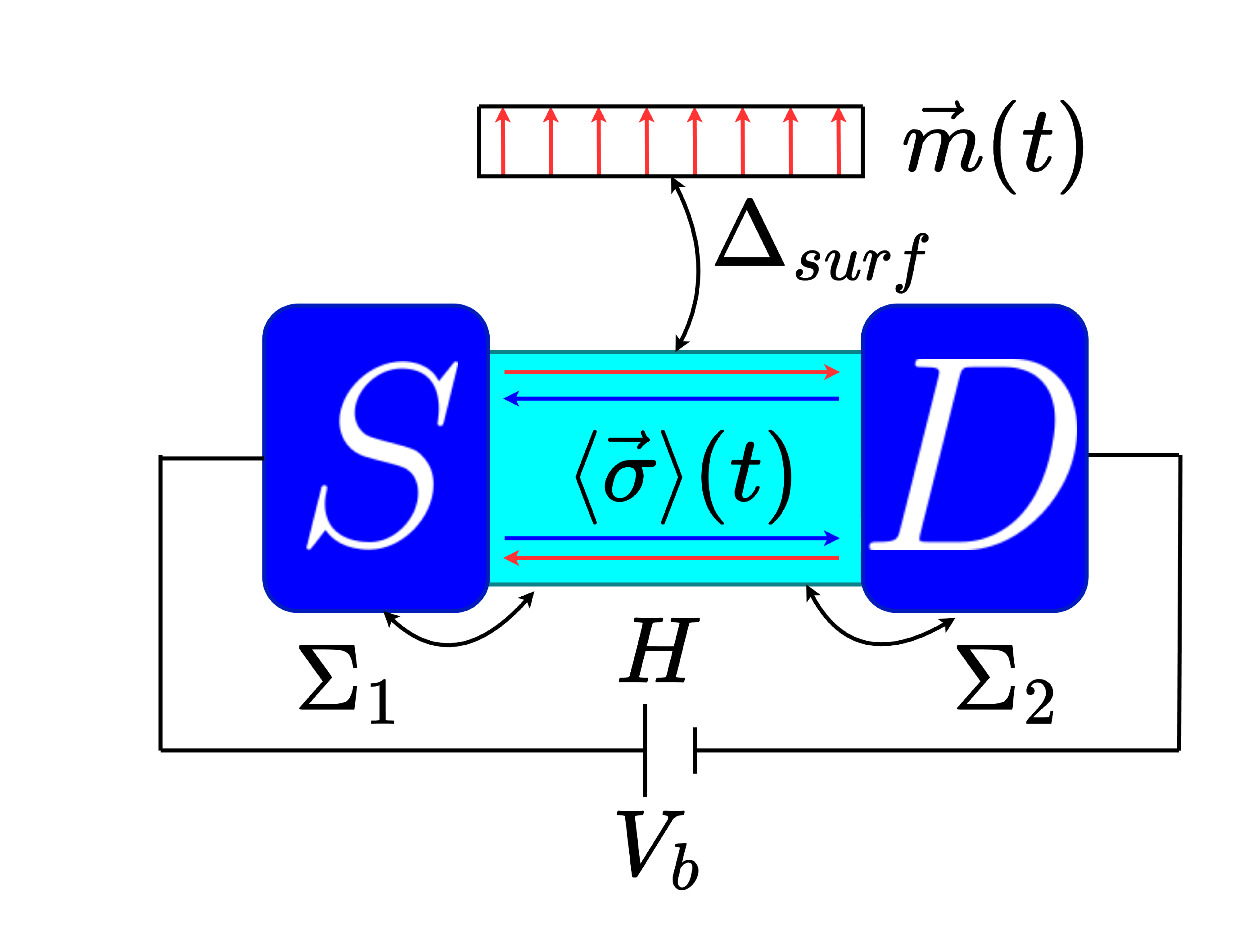}
    \caption{}
    \label{fig:Fig_6a}
    \end{subfigure}
    
    \vspace{0.5cm}

    \begin{subfigure}{0.8\textwidth}
    \centering
    \includegraphics[width=\textwidth]{plots/final_Figure_6b.pdf}
    \caption{}
    \label{fig:Fig_6b}
    \end{subfigure}
    
    \caption{Device schematic and flowchart of the methodology used. (a) Schematic of the two-terminal device. The channel consists of monolayer WTe$_2$ described by the Hamiltonian $H$, connected to contacts $\Sigma_{1,2}$ under an applied bias $V_b$. The monolayer is exchange-coupled to an adjacent magnetic layer with time-dependent magnetization $\vec{m}(t)$ via a surface exchange potential $\Delta_{\text{surf}}$. (b) Flowchart illustrating the simulation procedure for magnetization dynamics. At each time step, the exchange term in the Hamiltonian is calculated only over the lattice sites covered by the magnetic layer, which is then used within the non-equilibrium Green’s function (NEGF) formalism to compute the spin–orbit torque (SOT). The resulting torque is subsequently used in the Landau–Lifshitz equation to update the magnetization. This loop is repeated over the specified simulation time window.}
    \label{fig:Fig6ab}
\end{figure*}

In this appendix, we provide a detailed description of the theoretical framework used in our simulations. The formalism employed to compute spin--orbit torques (SOTs) follows \cite{Nikolic}. The device Hamiltonian is time dependent due to the dynamics of the magnetic moments in the ferromagnetic (FM) layer and is written as
\begin{equation}
\label{timedependentH}
\hat{H}(t) = \hat{H}_0 + \hat{U}_0(t),
\end{equation}
where $\hat{H}_0$ denotes the static part of the Hamiltonian and $\hat{U}_0(t)$ describes the exchange coupling between the device and the magnetic layer. Explicitly,
\begin{equation}
\hat{U}_0(t) = - \hat{c}_{i'}^{\dagger}
\left(
\frac{\Delta_{\mathrm{surf}}}{2}\,
\vec{m}(t)\cdot\vec{\sigma}
\right)
\hat{c}_{i'} ,
\end{equation}
where $\Delta_{\mathrm{surf}}$ is the interfacial exchange coupling strength, $\vec{m}_{i'}(t)$ is the unit magnetization vector at lattice site $i'$, and $\hat{\boldsymbol{\sigma}}$ denotes the Pauli matrices. Assuming an adiabatic approximation, valid when the magnetization dynamics is slow compared to electronic time scales, the spin density at the device-magnet interface can be expressed as
\begin{equation}
\label{spindensity}
s^i(t) \approx \int \frac{dE}{2\pi}
\left[
\sum_{\alpha} f_\alpha(E)\, T^{i\alpha}(E)
-
\frac{\Delta_{\mathrm{surf}}}{2}
\sum_j f'(E)\,
\frac{\partial m_j}{\partial t}\,
T^{ij}(E)
\right],
\end{equation}
where $f_\alpha(E)$ is the Fermi--Dirac distribution function of lead $\alpha$, $f'(E)$ is its energy derivative, $\alpha\in\{L,R\}$, and $i,j\in\{x,y,z\}$ label spin components. The energy-resolved transmission coefficients are given by the trace formulas
\begin{align}
\label{transmissiontrace}
T^{\alpha\beta}(E) &=
\mathrm{Tr}
\left[
\boldsymbol{\Gamma}_\alpha
\mathbf{G}_t
\boldsymbol{\Gamma}_\beta
\mathbf{G}_t^\dagger
\right], \\
T^{\alpha i}(E) &=
\mathrm{Tr}
\left[
\mathbf{1}_m
\boldsymbol{\sigma}_i
\mathbf{G}_t^\dagger
\boldsymbol{\Gamma}_\alpha
\mathbf{G}_t
\right],\\
T^{i \alpha}(E) &=
\mathrm{Tr}
\left[
\mathbf{1}_m
\boldsymbol{\sigma}_i
\mathbf{G}_t
\boldsymbol{\Gamma}_\alpha
\mathbf{G}_t^\dagger
\right], \\
T^{ij}(E) &=
\mathrm{Tr}
\left[
\mathbf{1}_m
\boldsymbol{\sigma}_i
(\mathbf{G}_t^\dagger-\mathbf{G}_t)
\mathbf{1}_m
\boldsymbol{\sigma}_j
(\mathbf{G}_t-\mathbf{G}_t^\dagger)
\right],
\end{align}
where $\boldsymbol{\Gamma}_\alpha$ denotes the coupling to lead $\alpha$, $\mathbf{G}_t$ is the retarded Green's function of the time-frozen Hamiltonian, and $\mathbf{1}_m$ projects onto the lattice sites covered by the magnetic layer. In the present device geometry, no current flows into the magnetic layer. Consequently, all torque contributions arise from the net spin density accumulated at the device--magnet interface. The magnetization dynamics is therefore governed by the Landau--Lifshitz equation
\begin{equation}
\label{LLGnew}
\frac{\partial \vec{m}}{\partial t}
=
- \frac{\Delta_{\mathrm{surf}}}{2}
\vec{m} \times \vec{s}(t).
\end{equation}

To identify different torque components, the transmission coefficients $T^{i\alpha}(E)$ are decomposed into parts that are even and odd under time-reversal symmetry:
\begin{align}
T_{\mathrm{e}}^{i\alpha}(E) &=
\frac{T^{i\alpha}(E) + T^{\alpha i}(E)}{2}, \\
T_{\mathrm{o}}^{i\alpha}(E) &=
\frac{T^{i\alpha}(E) - T^{\alpha i}(E)}{2}.
\end{align}
Each of these components can be further classified according to its symmetry under bias reversal:
\begin{align}
T_{\mathrm{e},\nu}^{i}(E)
&=
\frac{T_\nu^{iL}(E) + T_\nu^{iR}(E)}{2}, \\
T_{\mathrm{o},\nu}^{i}(E)
&=
\frac{T_\nu^{iL}(E) - T_\nu^{iR}(E)}{2}.
\end{align}
The physical significance of the resulting symmetry-resolved torque components is summarised in Table~\ref{tab:torque_symmetry}. For simplifying our simulations, we consider a constant anisotropy field strength.

\begin{table}
\centering
\begin{tabular}{|c|c|l|}
\hline\hline
$(\mu,\nu)$ & Symmetry & Physical interpretation \\
\hline
$(\mathrm{o},\mathrm{o})$ & Odd-odd   &Field-like torque along $(\vec{m}\times\vec{\sigma_P})$\\
$(\mathrm{e},\mathrm{e})$ & Even-even &Field-like torque along $(\vec{m}\times\vec{\sigma_P})$\\
$(\mathrm{o},\mathrm{e})$ & Odd-even  & Damping like torque along $\vec{m}\times(\vec{m}\times\vec{\sigma_P})$ \\
$(\mathrm{e},\mathrm{o})$ & Even-odd  & Vanishes identically \\
\hline\hline
\end{tabular}
\caption{Physical interpretation of the symmetry-resolved spin--orbit torque components. The indices $\mu$ and $\nu$ denote the symmetry of the torque under bias-reversal and time-reversal operations, respectively.}
\label{tab:torque_symmetry}
\end{table}

The corresponding energy-resolved spin--orbit torques are then given by
\begin{equation}
\label{resolvetorques}
\vec{\tau}_{\mu,\nu}(E)
=
- \frac{\Delta_{\mathrm{surf}}}{2}
\vec{m} \times \vec{T}_{\mu,\nu}(E),
\end{equation}
which are inserted into the Landau--Lifshitz equation to compute the magnetization dynamics. As illustrated in Fig.~\ref{fig:Fig6ab}b, the magnetization dynamics is obtained using a self-consistent iterative procedure. Starting from an initial magnetization configuration, the exchange term is constructed over the lattice sites covered by the magnetic layer and incorporated into the Hamiltonian. The resulting Hamiltonian is then used within the non-equilibrium Green's function formalism to compute the spin density and the associated spin-orbit torques. These torques are subsequently used to evolve the magnetization via the Landau--Lifshitz equation, yielding an updated magnetization configuration. This procedure is repeated until a steady state is reached or for a fixed simulation time window.
\twocolumngrid
\bibliography{apssamp}

@ARTICLE{Jia2017-tt,
  title     = "Direct visualization of a two-dimensional topological insulator
               in the {single-layer 1T$^\prime$-WTe$_2$}",
  author    = "Jia, Zhen-Yu and Song, Ye-Heng and Li, Xiang-Bing and Ran,
               Kejing and Lu, Pengchao and Zheng, Hui-Jun and Zhu, Xin-Yang and
               Shi, Zhi-Qiang and Sun, Jian and Wen, Jinsheng and Xing, Dingyu
               and Li, Shao-Chun",
  journal   = "Phys. Rev. B",
  publisher = "American Physical Society (APS)",
  volume    =  96,
  number    =  4,
  pages     =  {041108(R)},
  month     =  jul,
  year      =  2017,
  copyright = "http://link.aps.org/licenses/aps-default-license",
  doi       =  {https://doi.org/10.1103/PhysRevB.96.041108}
}

@ARTICLE{Duerloo2014-cw,
  title     = "Structural phase transitions in two-dimensional {Mo- and
               W}-dichalcogenide monolayers",
  author    = "Duerloo, Karel-Alexander N and Li, Yao and Reed, Evan J",
  journal   = "Nat. Commun.",
  publisher = "Springer Science and Business Media LLC",
  volume    =  5,
  number    =  1,
  pages     = "4214",
  month     =  jul,
  year      =  2014,
  copyright = "https://www.springernature.com/gp/researchers/text-and-data-mining",
  doi       =  {https://doi.org/10.1038/ncomms5214}
}

@ARTICLE{Isolating1Td,
  title     = "Water-assisted concerted layer growth of {T$_d$-phase WTe$_2$} for
               nonlinear {Hall} effect and microwave rectification",
  author    = "Wu, Shuang and Liu, Yu-Fei and Kwon, Sun Yong and Lim, Ting Yong
               and Yeom, Sinchul and Gao, Anyuan and Li, Houchen and Unocic,
               Raymond R and Varghese, Alina Mary and Lopez-Torres, Edgar and
               Xiao, Kai and Chang, Tay-Rong and Yoon, Mina and Xu, Su-Yang and
               Li, Xufan",
  journal   = "Nano Lett.",
  publisher = "American Chemical Society (ACS)",
  volume    =  25,
  number    =  50,
  pages     = "17475--17484",
  month     =  dec,
  year      =  2025,
  doi       =  {https://doi.org/10.1021/acs.nanolett.5c05085}
}

@ARTICLE{Kang2025-ed,
  title     = "Field-free spin-orbit torques switching and its applications",
  author    = "Kang, Min-Gu and Lee, Soogil and Park, Byong-Guk",
  journal   = "npj Spintronics",
  publisher = "Springer Science and Business Media LLC",
  volume    =  3,
  number    =  1,
  month     =  mar,
  year      =  2025,
  pages     =  "8",
  copyright = "https://creativecommons.org/licenses/by-nc-nd/4.0",
  doi       =  {https://doi.org/10.1038/s44306-025-00071-6}
}

@article{Qian_2014,
   title={Quantum spin {Hall} effect in two-dimensional transition metal dichalcogenides},
   volume={346},
   ISSN={1095-9203},
   url={http://dx.doi.org/10.1126/science.1256815},
   DOI={10.1126/science.1256815},
   number={6215},
   journal={Science},
   publisher={American Association for the Advancement of Science (AAAS)},
   author={Qian, Xiaofeng and Liu, Junwei and Fu, Liang and Li, Ju},
   year={2014},
   month=Dec, pages="1344–1347" }

@article{Wu_2018,
   title={Observation of the quantum spin {Hall} effect up to 100 kelvin in a monolayer crystal},
   volume={359},
   ISSN={1095-9203},
   url={http://dx.doi.org/10.1126/science.aan6003},
   DOI={10.1126/science.aan6003},
   number={6371},
   journal={Science},
   publisher={American Association for the Advancement of Science (AAAS)},
   author={Wu, Sanfeng and Fatemi, Valla and Gibson, Quinn D. and Watanabe, Kenji and Taniguchi, Takashi and Cava, Robert J. and Jarillo-Herrero, Pablo},
   year={2018},
   month=Jan, pages={76–79} }

@article{Fei_2017,
   title={Edge conduction in monolayer {WTe$_2$}},
   volume={13},
   ISSN={1745-2481},
   url={http://dx.doi.org/10.1038/nphys4091},
   DOI={10.1038/nphys4091},
   number={7},
   journal={Nature Physics},
   publisher={Springer Science and Business Media LLC},
   author={Fei, Zaiyao and Palomaki, Tauno and Wu, Sanfeng and Zhao, Wenjin and Cai, Xinghan and Sun, Bosong and Nguyen, Paul and Finney, Joseph and Xu, Xiaodong and Cobden, David H.},
   year={2017},
   month=Apr, pages={677–682} }

@article{SHE_rev,
  title = {Spin {H}all effects},
  author = {Sinova, Jairo and Valenzuela, Sergio O. and Wunderlich, J. and Back, C. H. and Jungwirth, T.},
  journal = {Rev. Mod. Phys.},
  volume = {87},
  issue = {4},
  pages = {1213--1260},
  numpages = {47},
  year = {2015},
  month = {Oct},
  publisher = {American Physical Society},
  doi = {10.1103/RevModPhys.87.1213},
  url = {https://link.aps.org/doi/10.1103/RevModPhys.87.1213}
}

@ARTICLE{vert_gradient,
  title     = "Field-free spin-orbit torque-induced switching of perpendicular
               magnetization in a ferrimagnetic layer with a vertical
               composition gradient",
  author    = "Zheng, Zhenyi and Zhang, Yue and Lopez-Dominguez, Victor and
               S{\'a}nchez-Tejerina, Luis and Shi, Jiacheng and Feng, Xueqiang
               and Chen, Lei and Wang, Zilu and Zhang, Zhizhong and Zhang, Kun
               and Hong, Bin and Xu, Yong and Zhang, Youguang and Carpentieri,
               Mario and Fert, Albert and Finocchio, Giovanni and Zhao,
               Weisheng and Khalili Amiri, Pedram",
  journal   = "Nature Communications",
  publisher = "Springer Science and Business Media LLC",
  volume    =  12,
  number    =  1,
  pages     = "4555",
  month     =  jul,
  year      =  2021,
  copyright = "https://creativecommons.org/licenses/by/4.0",
  doi       =  {https://doi.org/10.1038/s41467-021-24854-7}
}

@ARTICLE{interlayer_DMI,
  title     = "Field-free spin-orbit torque switching enabled by the interlayer
               {Dzyaloshinskii-Moriya} interaction",
  author    = "He, Wenqing and Wan, Caihua and Zheng, Cuixiu and Wang, Yizhan
               and Wang, Xiao and Ma, Tianyi and Wang, Yuqiang and Guo,
               Chenyang and Luo, Xuming and Stebliy, Maksim E and Yu, Guoqiang
               and Liu, Yaowen and Ognev, Alexey V and Samardak, Alexander S
               and Han, Xiufeng",
  journal   = "Nano Lett.",
  publisher = "American Chemical Society (ACS)",
  volume    =  22,
  number    =  17,
  pages     = "6857--6865",
  month     =  sep,
  year      =  2022,
  doi       =  {https://doi.org/10.1021/acs.nanolett.1c04786}
}

@ARTICLE{interlayer_ex_coupling,
  title     = "Spin-orbit torque switching without an external field using
               interlayer exchange coupling",
  author    = "Lau, Yong-Chang and Betto, Davide and Rode, Karsten and Coey, J
               M D and Stamenov, Plamen",
  journal   = "Nat. Nanotechnol.",
  publisher = "Springer Science and Business Media LLC",
  volume    =  11,
  number    =  9,
  pages     = "758--762",
  month     =  sep,
  year      =  2016,
  doi       =  {https://doi.org/10.1038/nnano.2016.84}
}

@ARTICLE{Liu2012-pv,
  title     = "Spin-torque switching with the giant spin {H}all effect of
               {T}antalum",
  author    = "Liu, Luqiao and Pai, Chi-Feng and Li, Y and Tseng, H W and
               Ralph, D C and Buhrman, R A",
  journal   = "Science",
  publisher = "American Association for the Advancement of Science (AAAS)",
  volume    =  336,
  number    =  6081,
  pages     = "555--558",
  month     =  may,
  year      =  2012,
  doi       =  {https://doi.org/10.1126/science.1218197}
}

@ARTICLE{Miron2011-sh,
  title     = "Perpendicular switching of a single ferromagnetic layer induced
               by in-plane current injection",
  author    = "Miron, Ioan Mihai and Garello, Kevin and Gaudin, Gilles and
               Zermatten, Pierre-Jean and Costache, Marius V and Auffret,
               St{\'e}phane and Bandiera, S{\'e}bastien and Rodmacq, Bernard
               and Schuhl, Alain and Gambardella, Pietro",
  journal   = "Nature",
  publisher = "Springer Science and Business Media LLC",
  volume    =  476,
  number    =  7359,
  pages     = "189--193",
  month     =  aug,
  year      =  2011,
  doi       =  {https://doi.org/10.1038/nature10309}
}

@ARTICLE{Worledge2024-ib,
  title={Spin-transfer torque magnetoresistive random access memory technology status and future directions},
  author={Daniel C. Worledge and Guo-Han Hu},
  journal={Nature Reviews Electrical Engineering},
  year={2024},
  volume={1},
  pages={730 - 747},
  url={https://api.semanticscholar.org/CorpusID:273893869},
  doi = {https://doi.org/10.1038/s44287-024-00111-z}
}

@ARTICLE{Berger1996-yb,
  title = {Emission of spin waves by a magnetic multilayer traversed by a current},
  author = {Berger, L.},
  journal = {Phys. Rev. B},
  volume = {54},
  issue = {13},
  pages = {9353--9358},
  numpages = {0},
  year = {1996},
  month = {Oct},
  publisher = {American Physical Society},
  doi = {10.1103/PhysRevB.54.9353},
  url = {https://link.aps.org/doi/10.1103/PhysRevB.54.9353}
}

@ARTICLE{STTFirst,
  title     = "Current-driven excitation of magnetic multilayers",
  author    = "Slonczewski, J C",
  journal   = "Journal of Magnetism and Magnetic Materials",
  publisher = "Elsevier BV",
  volume    =  159,
  number    = "1-2",
  pages     = "L1--L7",
  month     =  jun,
  year      =  1996,
  doi       =  {https://doi.org/10.1016/0304-8853(96)00062-5}
}

@ARTICLE{SOT_MTJ_rev,
   title={Spin-orbit torque switching of magnetic tunnel junctions for memory applications},
   volume={562},
   ISSN={0304-8853},
   url={http://dx.doi.org/10.1016/j.jmmm.2022.169692},
   DOI={10.1016/j.jmmm.2022.169692},
   journal={Journal of Magnetism and Magnetic Materials},
   publisher={Elsevier BV},
   author={Krizakova, Viola and Perumkunnil, Manu and Couet, Sébastien and Gambardella, Pietro and Garello, Kevin},
   year={2022},
   month=Nov, pages={169692} }

@ARTICLE{MRAMreview,
  title     = "Magnetoresistive random access memory",
  author    = "Apalkov, Dmytro and Dieny, Bernard and Slaughter, J M",
  journal   = "Proceedings of the IEEE",
  volume    =  104,
  number    =  10,
  pages     = "1796--1830",
  month     =  oct,
  year      =  2016,
  doi       = {https://doi.org/10.1109/JPROC.2016.2590142}
}

@ARTICLE{spintronics_MRAM_review,
  title     = "Spintronics based random access memory: a review",
  author    = "Bhatti, Sabpreet and Sbiaa, Rachid and Hirohata, Atsufumi and
               Ohno, Hideo and Fukami, Shunsuke and Piramanayagam, S N",
  journal   = "Mater. Today",
  publisher = "Elsevier BV",
  volume    =  20,
  number    =  9,
  pages     = "530--548",
  month     =  nov,
  year      =  2017,
  doi       = {http://dx.doi.org/10.1016/j.mattod.2017.07.007},
  copyright = "http://creativecommons.org/licenses/by-nc-nd/4.0/"
}

@ARTICLE{IEEE_symmetry,
  author={Zhou, Jing and Liu, Liang and Shu, Xinyu and Lin, Weinan and Zhao, Tieyang and Chen, Jingsheng},
  journal={IEEE Transactions on Magnetics}, 
  title={Symmetry Breaking by Materials Engineering for Spin-Orbit Torque Technology}, 
  year={2023},
  volume={59},
  number={8},
  pages={1-13},
  doi={10.1109/TMAG.2023.3283034}}

@article{1Td_theory,
  title = {Canted Persistent Spin Texture and Quantum Spin Hall Effect in {${\mathrm{WTe}}_{2}$}},
  author = {Garcia, Jose H. and Vila, Marc and Hsu, Chuang-Han and Waintal, Xavier and Pereira, Vitor M. and Roche, Stephan},
  journal = {Phys. Rev. Lett.},
  volume = {125},
  issue = {25},
  pages = {256603},
  numpages = {6},
  year = {2020},
  month = {Dec},
  publisher = {American Physical Society},
  doi = {10.1103/PhysRevLett.125.256603},
  url = {https://link.aps.org/doi/10.1103/PhysRevLett.125.256603}
}

@article{1Tp_akhmerov,
  title = {Influence of lattice termination on the edge states of the quantum spin {Hall} insulator monolayer {1T$^\prime$--WTe$_2$}},
  author = {Lau, Alexander and Ray, Rajyavardhan and Varjas, D\'aniel and Akhmerov, Anton R.},
  journal = {Phys. Rev. Mater.},
  volume = {3},
  issue = {5},
  pages = {054206},
  numpages = {9},
  year = {2019},
  month = {May},
  publisher = {American Physical Society},
  doi = {10.1103/PhysRevMaterials.3.054206},
  url = {https://link.aps.org/doi/10.1103/PhysRevMaterials.3.054206}
}

@article{1Td_experimental,
  title = {Determination of the Spin Axis in Quantum Spin {Hall} Insulator Candidate Monolayer {${\mathrm{WTe}}_{2}$}},
  author = {Zhao, Wenjin and Runburg, Elliott and Fei, Zaiyao and Mutch, Joshua and Malinowski, Paul and Sun, Bosong and Huang, Xiong and Pesin, Dmytro and Cui, Yong-Tao and Xu, Xiaodong and Chu, Jiun-Haw and Cobden, David H.},
  journal = {Phys. Rev. X},
  volume = {11},
  issue = {4},
  pages = {041034},
  numpages = {7},
  year = {2021},
  month = {Nov},
  publisher = {American Physical Society},
  doi = {10.1103/PhysRevX.11.041034},
  url = {https://link.aps.org/doi/10.1103/PhysRevX.11.041034}
}

@article{Nikolic,
  title = {Antidamping spin-orbit torque driven by spin-flip reflection mechanism on the surface of a topological insulator: A time-dependent nonequilibrium {Green} function approach},
  author = {Mahfouzi, Farzad and Nikoli\ifmmode \acute{c}\else \'{c}\fi{}, Branislav K. and Kioussis, Nicholas},
  journal = {Phys. Rev. B},
  volume = {93},
  issue = {11},
  pages = {115419},
  numpages = {10},
  year = {2016},
  month = {Mar},
  publisher = {American Physical Society},
  doi = {10.1103/PhysRevB.93.115419},
  url = {https://link.aps.org/doi/10.1103/PhysRevB.93.115419}
}

@article{Soleimani,
doi = {10.1209/0295-5075/117/37001},
url = {https://doi.org/10.1209/0295-5075/117/37001},
year = {2017},
month = {mar},
publisher = {EDP Sciences, IOP Publishing and Società Italiana di Fisica},
volume = {117},
number = {3},
pages = {37001},
author = {Soleimani, Maryam and Jalili, Seifollah and Mahfouzi, Farzad and Kioussis, Nicholas},
title = {Spin-orbit torque-driven magnetization switching in {2D}-topological insulator heterostructure},
journal = {Europhysics Letters},
}

@article{NEGF,
    author = "Keldysh, L. V.",
    title = "{Diagram Technique for Nonequilibrium Processes}",
    doi = "10.1142/9789811279461_0007",
    journal = "Sov. Phys. JETP",
    volume = "20",
    pages = "1018--1026",
    year = "1965"
}

@article{Manchon,
  title = {Current-induced spin-orbit torques in ferromagnetic and antiferromagnetic systems},
  author = {Manchon, A. and \ifmmode \check{Z}\else \v{Z}\fi{}elezn\'y, J. and Miron, I. M. and Jungwirth, T. and Sinova, J. and Thiaville, A. and Garello, K. and Gambardella, P.},
  journal = {Rev. Mod. Phys.},
  volume = {91},
  issue = {3},
  pages = {035004},
  numpages = {80},
  year = {2019},
  month = {Sep},
  publisher = {American Physical Society},
  doi = {10.1103/RevModPhys.91.035004},
  url = {https://link.aps.org/doi/10.1103/RevModPhys.91.035004}
}

@article{1Tp_experimental,
  author = {Shujie Tang and Chaofan Zhang and Dillon Wong and Zahra Pedramrazi and Hsin-Zon Tsai and Chunjing Jia and Brian Moritz and Martin Claassen and Hyejin Ryu and Salman Kahn and Juan Jiang and Hao Yan and Makoto Hashimoto and Donghui Lu and Robert G. Moore and Chan-Cuk Hwang and Choongyu Hwang and Zahid Hussain and Yulin Chen and Miguel M. Ugeda and Zhi Liu and Xiaoming Xie and Thomas P. Devereaux and Michael F. Crommie and Sung-Kwan Mo and Zhi-Xun Shen},
  title = {Quantum spin {Hall} state in monolayer {1T$^\prime$-WTe$_2$}},
  journal = {Nature Physics},
  volume = {13},
  number = {7},
  pages = {683--687},
  year = {2017},
  month = {Jul},
  doi = {10.1038/nphys4174},
  url = {https://doi.org/10.1038/nphys4174},
  issn = {1745-2481}
}

@ARTICLE{SOtvsSTT,
  title     = "Perspectives on field-free spin--orbit torque devices for memory
               and computing applications",
  author    = "Lopez-Dominguez, Victor and Shao, Yixin and Khalili Amiri,
               Pedram",
  journal   = "J. Appl. Phys.",
  publisher = "AIP Publishing",
  volume    =  133,
  number    =  4,
  pages     = "040902",
  month     =  jan,
  year      =  2023,
  doi       =  {https://doi.org/10.1063/5.0135185}
}

@misc{WTe2_1Tp_symmetry,
      title={Symmetry driven spin anisotropic magnetotransport in quantum spin {Hall} insulator {WTe$_2$ 1T}}, 
      author={Shrushti Tapar and Bent Weber and Saroj P Dash and Shantanu Mukherjee and Bhaskaran Muralidharan},
      year={2025},
      eprint={2507.18543},
      archivePrefix={arXiv},
      primaryClass={cond-mat.mes-hall}
}

@ARTICLE{Kwant,
  title     = "Kwant: a software package for quantum transport",
  author    = "Groth, Christoph W and Wimmer, Michael and Akhmerov, Anton R and
               Waintal, Xavier",
  journal   = "New J. Phys.",
  publisher = "IOP Publishing",
  volume    =  16,
  number    =  6,
  pages     = "063065",
  month     =  jun,
  year      =  2014,
  copyright = "http://creativecommons.org/licenses/by/3.0/",
  doi       =  {10.1088/1367-2630/16/6/063065}
}

@misc{STT_SOT_rev_Nikolic,
  title        = "First-principles quantum transport modeling of spin-transfer
                  and spin-orbit torques in magnetic multilayers",
  author       = "Nikolic, Branislav K and Dolui, Kapildeb and Petrovi{\'c},
                  Marko and Plech{\'a}{\v c}, Petr and Markussen, Troels and
                  Stokbro, Kurt",
  year         =  2018,
  primaryClass = "cond-mat.mes-hall",
  eprint       = "1801.05793",
  archivePrefix={arXiv},
}

@ARTICLE{STT_SOT_review2,
  author={Ikegawa, Sumio and Mancoff, Frederick B. and Janesky, Jason and Aggarwal, Sanjeev},
  journal={IEEE Transactions on Electron Devices}, 
  title={Magnetoresistive Random Access Memory: Present and Future}, 
  year={2020},
  volume={67},
  number={4},
  pages={1407-1419},
  doi={https://doi.org/10.1109/TED.2020.2965403}}

@ARTICLE{InPlaneBField1,
  title={Deterministic spin-orbit torque switching of epitaxial ferrimagnetic insulator with perpendicular magnetic anisotropy fabricated by on-axis magnetron sputtering},
  author={Roselle T. Ngaloy and Naoto Yamashita and Bing Zhao and Soojung Kim and Kohei Yamashita and Ivo P. C. Cools and M.N. Agusutrisno and Soobeom Lee and Yuichiro Kurokawa and Chun-Yeol You and Hiromi Yuasa and Saroj Prasad Dash},
  journal="npj Spintronics",
  volume=3,
  eid = 40,
  year=2025,
  doi = {https://doi.org/10.1038/s44306-025-00105-z}
}

@ARTICLE{another_SOT_reviw,
  title     = "Recent progress in spin-orbit torque magnetic random-access
               memory",
  author    = "Nguyen, V D and Rao, S and Wostyn, K and Couet, S",
  journal   = "npj Spintronics",
  publisher = "Springer Science and Business Media LLC",
  volume    =  2,
  number    =  1,
  pages     =  48,
  month     =  oct,
  year      =  2024,
  copyright = "https://creativecommons.org/licenses/by/4.0",
  doi       = {https://doi.org/10.1038/s44306-024-00044-1}
}

@ARTICLE{field_free_SOT_rev,
  title     = "Field-free approaches for deterministic spin-orbit torque
               switching of the perpendicular magnet",
  author    = "Wu, Hao and Zhang, Jing and Cui, Baoshan and Razavi, Seyed Armin
               and Che, Xiaoyu and Pan, Quanjun and Wu, Di and Yu, Guoqiang and
               Han, Xiufeng and Wang, Kang L",
  journal   = "Mater. Futures",
  publisher = "IOP Publishing",
  month     =  apr,
  year      =  2022,
  volume    =  1,
  eid    =  022201,
  doi       =  {10.1088/2752-5724/ac6577}
}

@ARTICLE{2DSOT,
  title     = "Two-dimensional materials for energy-efficient spin-orbit torque
               devices",
  author    = "Liu, Yuting and Shao, Qiming",
  journal   = "ACS Nano",
  publisher = "American Chemical Society (ACS)",
  volume    =  14,
  number    =  8,
  pages     = "9389--9407",
  month     =  aug,
  year      =  2020,
  doi       =  {https://doi.org/10.1021/acsnano.0c04403}
}

@ARTICLE{2H_phase_stability_claim,
  title     = "Phase engineering of transition metal dichalcogenides with
               unprecedentedly high phase purity, stability, and scalability
               via molten-metal-assisted intercalation",
  author    = "Park, Sanghyeon and Kim, Changmin and Park, Sung O and Oh, Nam
               Khen and Kim, Ungsoo and Lee, Junghyun and Seo, Jihyung and
               Yang, Yejin and Lim, Hyeong Yong and Kwak, Sang Kyu and Kim,
               Guntae and Park, Hyesung",
  journal   = "Adv. Mater.",
  publisher = "Wiley",
  volume    =  32,
  number    =  33,
  pages     = "2001889",
  month     =  aug,
  year      =  2020,
  copyright = "http://onlinelibrary.wiley.com/termsAndConditions\#vor",
  doi       =  {https://doi.org/10.1002/adma.202001889}
}

@ARTICLE{demo_SOT_WTe2,
  title     = "Control of spin--orbit torques through crystal symmetry in
               {WTe2/ferromagnet} bilayers",
  author    = "MacNeill, D and Stiehl, G M and Guimaraes, M H D and Buhrman, R
               A and Park, J and Ralph, D C",
  journal   = "Nature Physics",
  publisher = "Springer Science and Business Media LLC",
  volume    =  13,
  number    =  3,
  pages     = "300--305",
  month     =  mar,
  year      =  2017,
  doi       =  {https://doi.org/10.1038/nphys3933}
}

@ARTICLE{WTe2_1,
  title     = "Field-free magnetization switching induced by the unconventional
               spin--orbit torque from {WTe$_2$}",
  author    = "Xie, Qidong and Lin, Weinan and Sarkar, Soumya and Shu, Xinyu
               and Chen, Shaohai and Liu, Liang and Zhao, Tieyang and Zhou,
               Chenghang and Wang, Han and Zhou, Jing and Grade{\v c}ak,
               Silvija and Chen, Jingsheng",
  journal   = "APL Mater.",
  publisher = "AIP Publishing",
  volume    =  9,
  number    =  5,
  pages     = "051114",
  month     =  may,
  year      =  2021,
  copyright = "https://creativecommons.org/licenses/by/4.0/",
  doi       =  {https://doi.org/10.1063/5.0048926}
}

@ARTICLE{WTe2_2,
  title     = "Field-free spin--orbit torque-induced switching of perpendicular
               magnetization at room temperature in {WTe$_2$/ferromagnet}
               heterostructures",
  author    = "Wei, Lujun and Yin, Xiaomiao and Liu, Pai and Zhang, Pengchao
               and Niu, Wei and Liu, Ping and Yang, Jiajv and Peng, Jincheng
               and Huang, Fei and Liu, Ruobai and Chen, Jiarui and Chen, Lina
               and Zhou, Shuang and Li, Feng and Niu, Xianghong and Du, Jun and
               Pu, Yong",
  journal   = "Appl. Phys. Lett.",
  publisher = "AIP Publishing",
  volume    =  123,
  number    =  25,
  month     =  dec,
  year      =  2023,
  pages     =  252404,
  doi       =  {https://doi.org/10.1063/5.0048926}
}

@ARTICLE{WTe2_3,
  title     = "Deterministic switching of a perpendicularly polarized magnet
               using unconventional spin-orbit torques in {WTe$_2$}",
  author    = "Kao, I-Hsuan and Muzzio, Ryan and Zhang, Hantao and Zhu, Menglin
               and Gobbo, Jacob and Yuan, Sean and Weber, Daniel and Rao, Rahul
               and Li, Jiahan and Edgar, James H and Goldberger, Joshua E and
               Yan, Jiaqiang and Mandrus, David G and Hwang, Jinwoo and Cheng,
               Ran and Katoch, Jyoti and Singh, Simranjeet",
  journal   = "Nat. Mater.",
  publisher = "Springer Science and Business Media LLC",
  volume    =  21,
  number    =  9,
  pages     = "1029--1034",
  month     =  sep,
  year      =  2022,
  doi       =  {https://doi.org/10.1038/s41563-022-01275-5}
}

@ARTICLE{WTe2_4,
  title     = "Orbit-transfer torque driven field-free switching of
               perpendicular magnetization",
  author    = "Ye, Xing-Guo and Zhu, Peng-Fei and Xu, Wen-Zheng and Shang,
               Nianze and Liu, Kaihui and Liao, Zhi-Min",
  journal   = "Chin. Physics Lett.",
  publisher = "IOP Publishing",
  volume    =  39,
  number    =  3,
  pages     = "037303",
  month     =  feb,
  year      =  2022,
  copyright = "https://iopscience.iop.org/page/copyright",
  doi       =  {https://doi.org/10.1088/0256-307X/39/3/037303}
}

@ARTICLE{WTe2_5,
  title     = "Field-free deterministic switching of all-{van der Waals}
               spin-orbit torque system above room temperature",
  author    = "Kajale, Shivam N and Nguyen, Thanh and Hung, Nguyen Tuan and Li,
               Mingda and Sarkar, Deblina",
  journal   = "Sci. Adv.",
  publisher = "American Association for the Advancement of Science (AAAS)",
  volume    =  10,
  number    =  11,
  pages     = "eadk8669",
  month     =  mar,
  year      =  2024,
  doi       =  {https://doi.org/10.1126/sciadv.adk8669}
}

@ARTICLE{WTe2_6,
  title     = "Cascadable in-memory computing based on symmetric writing and
               readout",
  author    = "Wang, Lizheng and Xiong, Junlin and Cheng, Bin and Dai, Yudi and
               Wang, Fuyi and Pan, Chen and Cao, Tianjun and Liu, Xiaowei and
               Wang, Pengfei and Chen, Moyu and Yan, Shengnan and Liu, Zenglin
               and Xiao, Jingjing and Xu, Xianghan and Wang, Zhenlin and Shi,
               Youguo and Cheong, Sang-Wook and Zhang, Haijun and Liang,
               Shi-Jun and Miao, Feng",
  journal   = "Sci. Adv.",
  publisher = "American Association for the Advancement of Science (AAAS)",
  volume    =  8,
  number    =  49,
  pages     = "eabq6833",
  month     =  dec,
  year      =  2022,
  doi       =  {https://doi.org/10.1126/sciadv.abq6833}
}

@ARTICLE{SP1,
  title     = "Tunable unconventional spin orbit torque magnetization dynamics
               in van der Waals heterostructures",
  author    = "Pandey, Lalit and Zhao, Bing and Tenzin, Karma and Ngaloy,
               Roselle and Lamparsk{\'a}, Veronika and Bangar, Himanshu and
               Ali, Aya and Abdel-Hafiez, Mahmoud and Zhang, Gaojie and Wu, Hao
               and Chang, Haixin and Sj{\"o}str{\"o}m, Lars and Rout, Prasanna
               and S{\l}awi{\'n}ska, Jagoda and Dash, Saroj P",
  journal   = "Nature Communications",
  publisher = "Springer Science and Business Media LLC",
  volume    =  16,
  number    =  1,
  pages     = "8722",
  month     =  sep,
  year      =  2025, 
  doi       = "https://doi.org/10.1038/s41467-025-64109-3"
}

@ARTICLE{SP2,
  title     = "Nanoelectronics with two-dimensional magnets",
  author    = "Zhao, Bing and Ngaloy, Roselle and Pandey, Lalit and Bangar,
               Himanshu and Dubey, Divya P and Dash, Saroj P",
  publisher = "American Chemical Society (ACS)",
  journal   = "Nano Letters",
  volume    =  26,
  number    =  22,
  pages     = "7209--7226",
  month     =  jun,
  year      =  2026,
  doi       = "https://doi.org/10.1021/acs.nanolett.6c00983"
}

@ARTICLE{StrainInWTe2,
  title     = "Realistic tight-binding model for monolayer transition metal
               dichalcogenides of 1{T}$^\prime$ structure",
  author    = "Hu, Mengli and Ma, Guofu and Wan, Chun Yu and Liu, Junwei",
  journal   = "Phys. Rev. B",
  publisher = "American Physical Society (APS)",
  volume    =  104,
  pages     =  035156,
  month     =  jul,
  year      =  2021,
  copyright = "https://link.aps.org/licenses/aps-default-license",
  doi       = "https://doi.org/10.1103/PhysRevB.104.035156 "
}

\end{document}